\documentclass[final,5p,times,twocolumn]{elsarticle} 

\usepackage{lineno,hyperref}
\modulolinenumbers[5]
\usepackage{siunitx}
\usepackage{caption}
\usepackage{amsmath}
\journal{Nuclear Instruments \& Methods in Physics Research, Section A}
\makeatletter
\renewcommand\paragraph{%
  \@startsection{paragraph}{4}{\z@}%
    {6pt}               
    {-0.5em}            
    {\normalfont\normalsize\itshape}%
}%
\makeatother

\usepackage{numcompress}
\begin{document}

\begin{frontmatter}

\title{Annihilation Vertex Reconstruction Algorithm with Single-Layer Timepix4 Detectors}

\author[a,b]{V.~Kraxberger\corref{mycorrespondingauthor}}
\cortext[mycorrespondingauthor]{Corresponding author}
\ead{viktoria.kraxberger@oeaw.ac.at}

\author[e]{A.~Gligorova}
\author[a]{E.~Wid\-mann}




\address[a]{Stefan Meyer Institute, Austrian Academy of Science, Dominikanerbastei 16, Vienna, Austria}
\address[b]{Vienna Doctoral School in Physics, University of Vienna, Boltzmanngasse 5, Vienna, Austria}
\address[e]{Faculty of Physics, University of Vienna, Boltzmanngasse 5, Vienna, Austria}

\begin{abstract}
A study of antiproton-nucleus annihilations at rest on a variety of thin solid targets using slow extracted antiprotons is being prepared. 
To detect the charged annihilation products, the experiment will employ seven Timepix4 ASICs coupled to \qty{500}{\um} thick silicon sensors. These will be arranged in a cuboid geometry that covers the majority of the full solid angle around the target, enabling precise tracking of outgoing particles using only one layer of detectors. With these novel chips, the annihilation will be studied by measuring the total multiplicity, energy, and angular distribution of various prongs produced in a number of targets.\\
A 3D reconstruction algorithm for determining the annihilation vertex from particle tracks in the single-plane detectors has been developed using Monte Carlo simulations. This allows for event-by-event reconstruction, making it possible to distinguish antiproton annihilations on the target from those occurring elsewhere. 
The measurements will also enable a study of possible final state interactions triggered by the primary annihilation mesons, their evolution with the nuclear mass and their branching ratios.
\end{abstract}

\begin{keyword}
Vertex reconstruction\sep Semiconductor detector \sep Pixel detector\sep Antimatter physics \\
Vienna Conference on Instrumentation 2025
\end{keyword}

\end{frontmatter}


\section{Introduction}
Most antimatter experiments aim to precisely measure its properties in search for structural differences to the matter counterparts. Since the detection of antimatter is mainly done through its annihilation \cite{Sauerzopf2017,Nagata2018,Storey2013,Andresen2012,Capra2017,Aumann2022}, the understanding of the antiproton-nucleus (\=pA) interaction at low energies is crucial. Previous measurements of \=p annihilations showed that current Monte Carlo simulation models do not reproduce the measured data precisely \citep{Amsler2024, Aghion2017}, as the annihilation mechanism, in terms of participating nucleons and their constituents, is not well established. \\
This project is a detailed study of the production of charged fragments (prongs) in antiproton-induced reactions across a range of nuclei at keV incident energy. The main goal is to conduct an in-depth investigation of the ﬁnal state interactions (FSIs), triggered by primary annihilation mesons \cite{Kraxberger2025a}.

\paragraph{Annihilation Process.} An antiproton at rest is captured by the atom and cascades down until it reaches an orbit close enough to the nuclear surface for it to annihilate on a nucleon \cite{Plendl1993}. The initial products of the annihilation are mesons. On average, for annihilation on a single proton, five pions are produced of which three are charged \cite{Klempt2005, Amsler1998}. These pions, which have average kinetic energies of about \qty{235}{\MeV}, can penetrate the nucleus, excite it, and lead to the emission of protons, neutrons or heavier fragments \cite{Plendl1993}. This vast variety of possible FSIs has not yet been completely described by the existing models. 
Branching ratios have only been reported for annihilations on He nuclei \cite{Montagna2002, Bendiscioli1990} and the total prong multiplicities for a very limited number of nuclei \cite{Bendiscioli1994}.

\paragraph{Detector Requirements.} A detection system designed for the study of \=pA annihilations must be capable of measuring both the primarily produced mesons, which are minimum ionizing particles (MIPs), and heavier fragments, which are heavily ionising particles (HIPs). Given that the annihilation products are expected to be isotropically distributed, a detector with large solid-angle coverage is advantageous. To determine the MIP and HIP multiplicities for each annihilation, precise event tagging is required. Consequently, the detector must have sufficient time resolution to reliably distinguish individual annihilating antiprotons and correctly associate them with their respective prongs.

\paragraph{Antiproton Beam.} This experiment requires a focussed antiproton beam with energies that allow them to stop within the 1-\qty{3}{\um} thick target foils. This is compatible with the trapping potentials used in experiments at CERN's Antimatter Factory, typically of \qty{10}{\keV}. To resolve individual annihilation events, the rate must stay below the Timepix4 detector’s per-pixel dead time ($\sim \qty{500}{\ns}$), necessitating a slowly extracted beam.

\section{Detection System and Target}
\begin{figure}[t]
    \centering
    \includegraphics[width=0.9\linewidth]{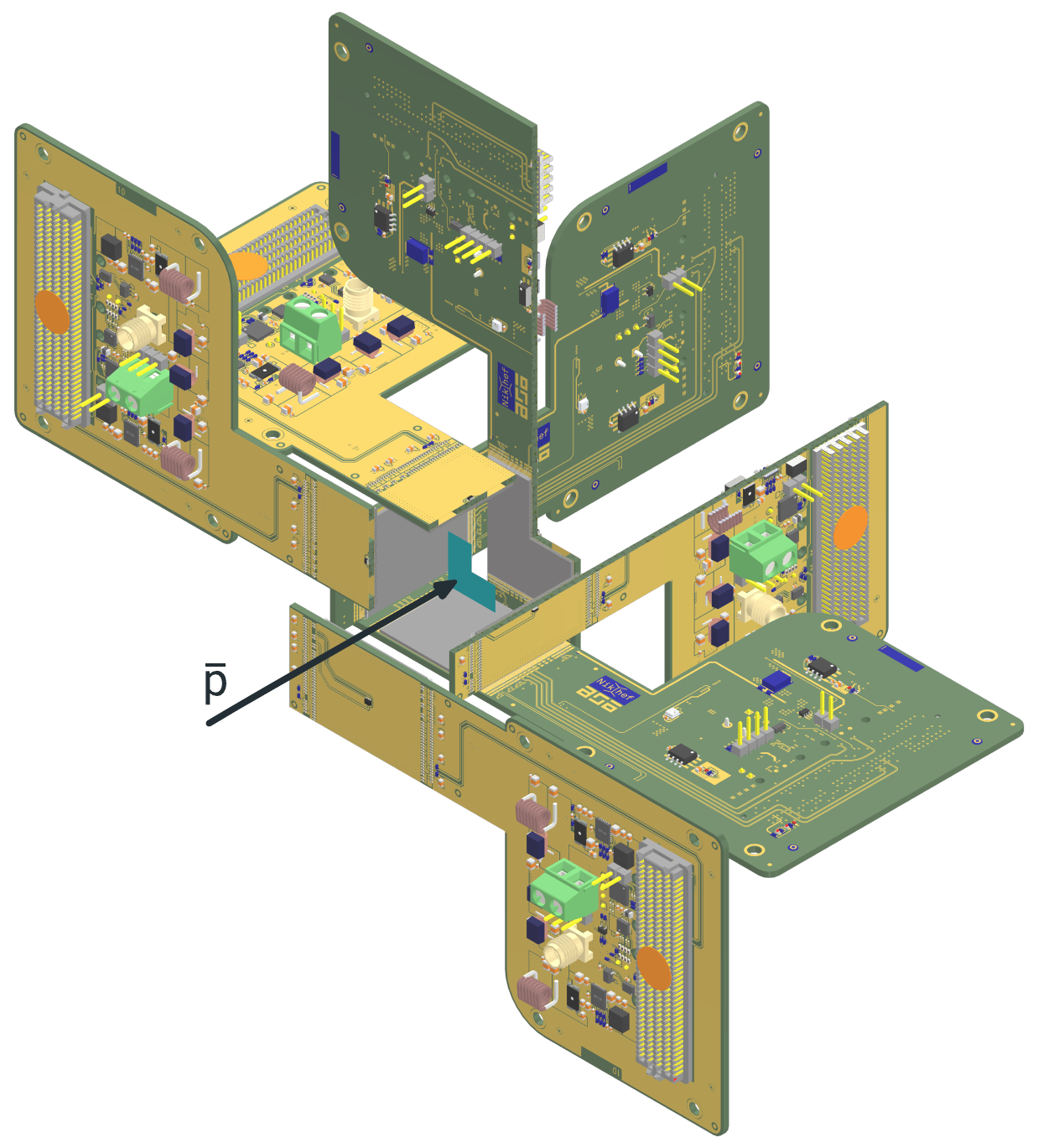}
    \vspace*{-1ex}
    \caption{Three-quarter section of the seven Timepix4 chips on their L-shaped carrier boards. The target foil in the centre is shown in blue and the direction of the incoming antiproton beam is indicated by an arrow.}
    \label{fig:board_geometry}
\end{figure}

The detector consists of seven Timepix4 hybrid pixel readout chips \cite{Llopart2022} coupled to \qty{500}{\um} silicon sensors, arranged in a cuboid geometry. With this configuration the majority of the solid angle around a \qtyproduct[product-units = bracket-power]{1 x 1}{\cm} target foil is covered. The ASICs (application specific integrated circuits) have 448 $\times$ 512 pixels with a \qty{55}{\um} pitch. The area of one sensor is \qtyproduct[product-units = bracket-power]{24.64 x 28.16}{\mm}. L-shaped chip carrier boards for the Timepix4 were specifically designed and provided by the NIKHEF institute, enabling the tight placement of the detectors, see Fig. \ref{fig:board_geometry}. The bin size for the time-of-arrival (ToA) of the signals is below \qty{200}{\ps} and the energy resolution of the time-over-threshold (ToT) is $1$-$\qty{2}{\keV}$. 

\section{Monte Carlo Simulation}
Simulations were carried out in both Geant4 \cite{Allison2016, Allison2006, Agostinelli2003} and FLUKA \cite{Ahdida2022, Battistoni2015}. The geometry of the setup includes the target foil, its mounting structure, the chip carrier boards and the vacuum chamber itself. 
In total 12 different foils were simulated, ranging from carbon to lead, as well as antiproton beams between \qty{250}{\eV} and \qty{10}{\keV} kinetic energy. Several physics models in Geant4 were explored, but ultimately the list FTFP\_INCLXX\_EMZ (v11.2.1) was used,
which includes the Intranuclear Cascade de Liège (INCL) model \cite{Boudard2013} for antiproton annihilations at rest \cite{Zharenov2023}. To obtain sufficient information about the process within the target and sensors, the step size of a particle propagation was chosen to be \qty{\leq 5}{\um}.\\
The simulations from both models were saved as ROOT \cite{ROOT} files, which were then digitised with the Allpix$^2$ Semiconductor Detector Monte Carlo Simulation Framework \cite{Spannagel2018}, setting a bias voltage of \qty{150}{\V} and a depletion voltage of \qty{100}{\V} for the p-on-n sensor. 

\section{Vertex Reconstruction Algorithm}
\begin{table*}[t]
    \centering
    \begin{tabular}{c}
        \includegraphics[width=\textwidth]{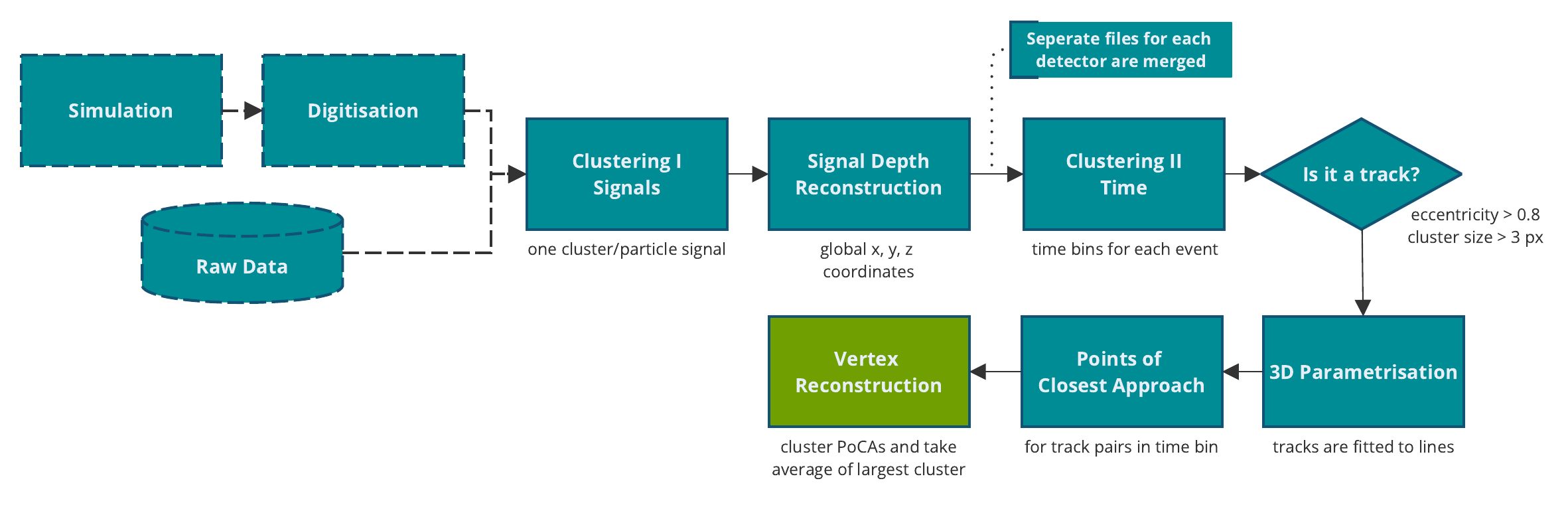} \\
    \end{tabular}
    \vspace*{-3ex}
    \captionof{figure}{Flowchart showing the vertex reconstruction process. The same process can be applied to both simulated data and raw data from real measurements.}
    \label{fig:flow}
\end{table*}

\begin{figure}[b!]
    \centering
    \makebox[\textwidth]{
    \resizebox{\textwidth}{!}{\includegraphics[width=\textwidth]{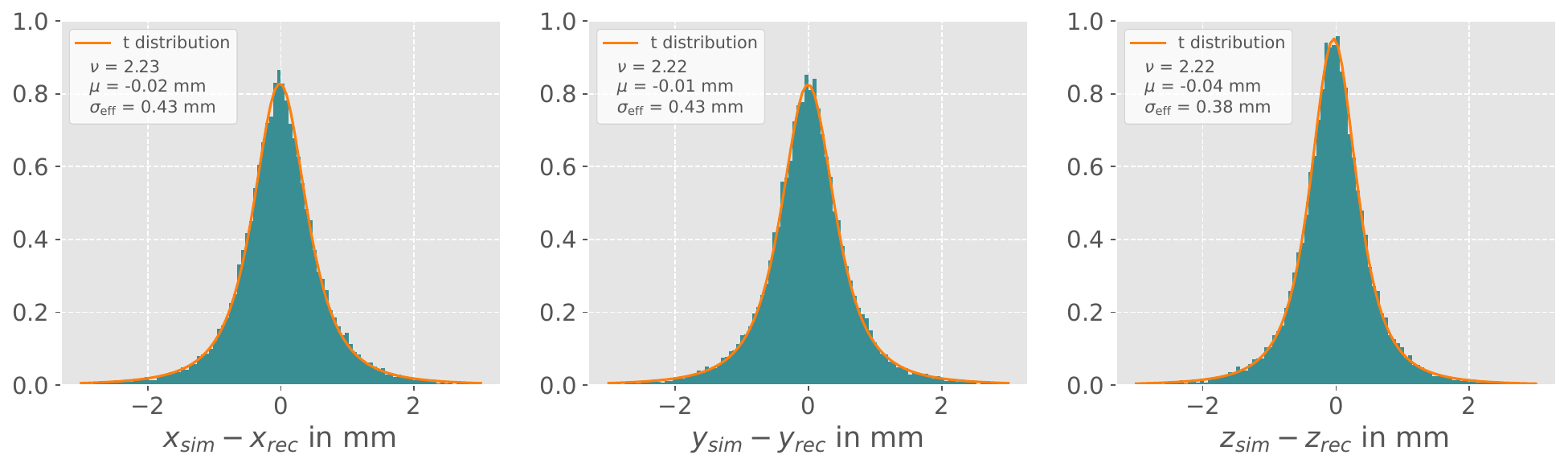}}}
    \vspace*{-2ex}
    \captionsetup{justification=justified, width=\textwidth, margin=0cm}
    \makebox[\textwidth]{%
        \parbox{\textwidth}{%
        \caption{The differences between simulated and reconstructed vertex positions for $x, y, z$, fitted with Student-t distributions.}
        \label{fig:xyz_diff}
        }
    }
\end{figure}

The data acquisition system of the Timepix4 provides a zero-suppressed, constant stream. To tag individual events in the data, the annihilation vertex is reconstructed for events with two or more tracks in the detectors. The algorithm has been developed using the simulations described in the previous section. In Fig. \ref{fig:flow} the whole process is shown as a flowchart and each step will be described here.
\paragraph{Clustering I, Signals.}
Each detector produces its own data file, thus the Allpix$^2$ output is also saved as seven individual files and the same code can be applied to both simulation and raw data. As a first step, the pixel signals are clustered such that all pixel hits belonging to a single particle are grouped. For the clustering, the DBSCAN \cite{Schubert2017} algorithm from python's Scikit-learn libary \cite{ScikitLearn} is applied to an array of pixel columns, rows and the ToA information. This algorithm was chosen as the number of clusters is not initially known and their shapes can vary from straight lines (MIPs), curls (electrons) or large blobs (HIPs). After this, each signal is assigned a cluster ID, the total number of cluster IDs corresponds to the total number of particles measured.
\paragraph{Signal Depth Reconstruction.}
The position of a signal is obtained from the pixel column and row, which can then easily be converted to $x$ and $y$ coordinates by knowing the geometry. In order to get the third coordinate $z$ within the \qty{500}{\um} silicon sensor, the depth of the signal has to be reconstructed. This is possible with several methods \cite{Manek2022}. One method relies on the lowest and highest ToA within a cluster, assigning them as entry and exit points of a particle track, given the particle is not stopped within the sensor. The method applied here makes use of the known drift time of the charge carries within the sensor. The local $z$ coordinate can be calculated using the $\Delta\mathrm{ToA}$ within a cluster, the bias voltage $U_B$ applied to the sensor, the depletion voltage $U_D$, the mobility of the charge carriers (in this case holes) $\mu_h$ and the thickness of the sensor $d$ \cite{Bergmann2017}:
\begin{equation}
    z (\Delta\mathrm{ToA}) = \frac{d}{2 U_D} (U_D + U_B) \left( 1 - \exp \left( \frac{2 U_D \mu_h t}{d^2}  \right) \right)
    \label{equ:z_rec}
\end{equation}
This way, the local coordinates of pixel column, row and ToA are converted to global $x, y, z$ coordinates, where $(0,0,0)$ is the centre of the target foil.
\paragraph{Clustering II, Time.}
In this step the individual files for the seven detectors are merged to one, which then undergoes a second clustering process using the first time stamp of each cluster. This way all clusters (particles) belonging to one event (annihilation) are grouped together.
\paragraph{Is it a track?}
Each cluster can be assumed to have an elliptical shape for which the eccentricity can be calculated as focal distance over major axis length. As HIPs are mostly stopped within the sensor and deposit all their energy, they leave rounder clusters. Elongated straight clusters - tracks - are mostly produced by MIPs traversing the sensor. These tracks are crucial for the vertex reconstruction and therefore the whole data set is filtered for clusters with eccentricity $ > 0.8$ and a cluster size $ > 3$~pixels.
\paragraph{3D Parametrisation and Points of Closest Approach.}
Each track can be fitted by a 3D linear function using singular value decomposition, resulting in a parametrisation with one point on the line, its average, and a directional vector. Using these parameters the point of closest approach (PoCA) can be calculated. This is done for all possible pairs of $n$ tracks within each time bin, resulting in ${n \choose 2}$ points per event. Again DBSCAN is used on the PoCAs in each time bin, as this algorithm can also exclude outliers which could arise from cosmic rays or secondary particles decaying elsewhere. The parameter $\varepsilon$ (maximum distance between two points for them to be considered as neighbours) was set to \qty{5}{\mm}, after exploring its effect on the number of reconstructed events. For $\varepsilon > \qty{5}{\mm}$, this number showed no significant change. The final value for the reconstructed vertex $\vec{V}_\mathrm{rec}$ is taken as the average point of the largest cluster of PoCAs.\\
The simulated point of annihilation $\vec{V}_\mathrm{sim}$ are compared to $\vec{V}_\mathrm{rec}$ in each coordinate, see Fig. \ref{fig:xyz_diff}. The histograms were fitted with Student-t distributions, as the tails are heavier than in normal distributions. This can be explained by the dependence of the resolution on the number of prongs measured. The means $\mu$ of the distributions are very close to zero, showing there is no bias in the vertex fitting. The scale parameter $\sigma_\mathrm{eff}$ and the degrees of freedom $\nu$ can be used to approximate the standard deviation of the distribution as $\sigma = \sigma_\mathrm{eff} \sqrt{\nu / (\nu - 2)}$:

{\centering
  $ \displaystyle
    \begin{aligned} 
       \sigma_x = \qty{1.34}{\mm} \qquad \sigma_y = \qty{1.36}{\mm} \qquad \sigma_z = \qty{1.21}{\mm}
    \end{aligned}
  $ 
\par}




\section{Summary and Outlook}
An algorithm has been developed through simulations, capable of reconstructing antiproton annihilation vertices using a single-layer silicon detector. It accurately identifies individual annihilation events on different nuclei, offering a spatial resolution below 1.5 mm, sufficient to distinguish between annihilations occurring on the target and those elsewhere. 
This work will be applied to a new \=pA annihilation experiment at CERN’s Antimatter Factory, focusing on prong measurements for various target foils, and enabling further investigation of the possible final state interactions and their evolution with nuclear mass.

\section{Acknowledgments}
The authors express their gratitude to the Medipix4 Collaboration at CERN for providing the Timepix4 ASICs for the upcoming measurements, NIKHEF for providing the custom designed chip carrier boards, in particular to Martin van Beuzekom, and to the late Hannes Zmeskal for his substantial contribution in the design of the experiment.
This research was funded in whole or in part by the Austrian Science Fund (FWF) P 34438. For open access purposes, the author has applied a CC BY public copyright license to any author accepted manuscript version arising from this submission.

\newpage
\bibliography{references}

\end{document}